\documentclass[conference,10pt,letter]{IEEEtran}
\usepackage{graphicx}
\usepackage{amsmath}
\usepackage{amssymb}
\usepackage{definice}
\usepackage{calc}
\usepackage{multirow}
\usepackage{mathrsfs}
\usepackage[english]{babel}
\usepackage{float}
\usepackage{cool}
\IEEEoverridecommandlockouts
\usepackage{times}
\usepackage{color,colortbl}
\usepackage{tikz}
\usepackage{caption}
\usepackage{subcaption}
\usetikzlibrary{calc,arrows,positioning}
\definecolor{TableCol}{rgb}{1,0.75,0}
\definecolor{TableCol2}{rgb}{1,0.9,0}
\definecolor{TableCol3}{rgb}{1,0.5,0}
\usepackage{enumitem}
\usepackage{commath}
\usepackage{tensor}
\usepackage{mathtools}
\usepackage[thref]{ntheorem}
\usepackage{textcomp}
\usepackage{tabstackengine}
\usepackage[makeroom]{cancel}

\usepackage{textcomp}

\usepackage{times}
\usepackage{amssymb}

\usepackage{cite} 
\usepackage{stfloats}  

\newcounter{steps}
	{\end{list}}

\def\bfA{\mathbf A}

\def\bfC{\mathbf C}
\def\bfD{\mathbf D}
\def\bfF{\mathbf F}
\def\bfG{\mathbf G}

\def\bfI{\mathbf I}

\def\bfK{\mathbf K}

\def\bfQ{\mathbf Q}
\def\bfP{\mathbf P}
\def\bfR{\mathbf R}
\def\bfS{\mathbf S}

\def\bfg{\mathbf g}
\def\bfh{\mathbf h}

\def\bfu{\mathbf u}
\def\bfv{\mathbf v}
\def\bfw{\mathbf w}
\def\bfx{\mathbf x}
\def\bfy{\mathbf y}
\def\bfz{\mathbf z}

\def\mean{\mathsf{E}}
\def\var{\mathsf{var}}
\def\cov{\mathsf{cov}}

\def\bfxi{\ensuremath{\boldsymbol{\xi}}}

\DeclareMathAlphabet\mathbfcal{OMS}{cmsy}{b}{n}

\begin{document}
	
	\title{Design of Efficient Point-Mass Filter with Terrain Aided Navigation Illustration}
	

\author{J. Matou\v{s}ek, J. Dun\'{i}k, M. Brandner 
	\thanks{Authors are with Department of Cybernetics, Pilsen, Czech Republic. E-mails: e-mail: \{dunikj,matoujak\}@kky.zcu.cz (J. Dun\'{i}k, J. Matou\v{s}ek), and Department of Mathematics, Faculty of Applied Sciences, University of West Bohemia, brandner@kma.zcu.cz (M. Brandner).}
}
	
	\maketitle
	
	\selectlanguage{english}
	\begin{abstract}
		\noindent This paper deals with state estimation of stochastic models with linear state dynamics, continuous or discrete in time. The emphasis is laid on a numerical solution to the state prediction by the time-update step of the grid-point-based point-mass filter (PMF), which is the most computationally demanding part of the PMF algorithm. A novel efficient PMF (ePMF) estimator, unifying continuous and discrete, approaches is proposed, designed, and discussed. By numerical illustrations, it is shown, that the proposed ePMF can lead to a time complexity reduction that exceeds $\mathbf{99.9\%}$ without compromising accuracy. The MATLAB\textregistered\ code of the ePMF is released with this paper.
	\end{abstract}

\begin{IEEEkeywords}
State estimation, transition probability matrix, Chapman-Kolmogorov equation, Fokker-Planck equation, point-mass filter, convolution, terrain-aided navigation.
\end{IEEEkeywords}

	\section{Introduction}
	State estimation of discrete-time stochastic dynamic systems from noisy measurements has been a subject of considerable research interest for the last decades. Following the Bayesian approach, a general solution to the state estimation problem is given by the Bayesian recursive relations (BRRs) computing the probability density functions (PDFs) of the state conditioned on the measurements. The conditional PDFs provide a full description of the immeasurable state of a nonlinear or non-Gaussian stochastic dynamic system. The relations are, however, exactly tractable for a limited set of models for which linearity and Gaussianity is usually common factor. This class of exact Bayesian estimators is represented e.g., by the Kalman filter (KF). In other cases, an approximate solution to the BRRs has to be employed. These approximate filtering methods can be divided with respect to the validity of the estimates into global and local filters \cite{So:74}.

The \textit{local} filters provide computationally efficient estimates predominantly in the form of the conditional mean and covariance matrix\footnote{First two moments usually do not represent a full description of the immeasurable state.} with potentially limited performance due to the inherent underlying Gaussian assumption, which is not always realistic. These filters are represented by the extended Kalman filter, the unscented Kalman filter, the cubature Kalman filter, or the stochastic integration filter \cite{AnMo:79,JuUhl:04,DuStrSi:13,JiXiChe:13}. 

As opposed to the local filters, the \textit{global} filters provide estimates in the form of conditional PDFs. The global filters are capable of estimating the state of a strongly nonlinear or non-Gaussian system but at the cost of higher computational demands. Among these, the Gaussian sum filter \cite{SoAl:71}, the particle filter (PF) \cite{DoFrGo-book:01}, and the point-mass filter (PMF) \cite{Be:99,SiKraSo:02,KaSch13,Ma:20} have attracted a considerable attention. 

This paper considers the point-mass filter. The PMF, developed in the seventies \cite{So:74}, evaluates the conditional PDFs only at the grid of points covering a significant part of the continuous state space. It is based on the numerical integration in the discrete model dynamics case and the numerical method solution in the case of the continuous dynamics. This leads to a considerably computationally complex estimator. The main computational load is located in the time-update, where the probability of each filtering grid point moving to each predictive grid point has to be calculated in the discrete case, and the numerical method is solved in the continuous case. Therefore, the PMF is considered the more computationally expensive estimator in comparison with its main competitor PF. On the other hand, its results are deterministic and it is more robust.

In our previous paper \cite{MaDuBrPaCh:23} we have focused on the reduction of the computational complexity of the time-update and proposed an efficient point-mass predictor, (ePMFp) which simplified the big $\mathbfcal{O}$ complexity from $\mathbfcal{O}(N^2)$ to $\mathbfcal{O}(N \log (N))$ for systems with linear dynamics, where $N$ is the number of grid points.

The goal of this paper is to extend the efficient point-mass predictor to efficient PMF (ePMF) for both continuous and discrete dynamics cases, look more thoroughly into the estimation of the continuous dynamics, apply continuous and discrete resulting estimation algorithms to terrain-aided navigation scenarios, and compare them to the PF. The ePMF is detailed with the stress laid on solving implementation issues. Alongside this paper, ePMF TAN (terrain-aided navigation) MATLAB\textregistered\ implementations for a random walk (2 dimensions) and coordinated turn dynamics (4 dimensions) are released. Most importantly, the implementations show, that the efficient version of the PMF can now compete with PF in both accuracy and computational complexity.

The rest of this paper is organized as follows. In Section II, used TAN models are defined, and state estimation is introduced. Section III presents the standard PMF algorithm. The idea of the novel efficient point-mass predictor is laid out in Section IV. Follows the proposal of the ePMF in Section V. Section VI describes the enclosed MATLAB\textregistered\ implementation. The numerical results are presented in Section VII. And the paper ends with Section VIII, which is containing concluding remarks.

		\section{Terrain Aided Navigation and State Estimation} \label{sec:tan}
The terrain-aided navigation utilizes a measurement of a terrain altitude that is below the vehicle. The terrain altitude can be measured by various altimeters. This measurement is then correlated with the horizontal positions via the map, to correct the predicted position. Alternatively, instead of height measurement correlation with terrain map, a gravity measurement can be correlated with gravity map on the same principle \cite{MuBrBiZa:17}. The base for all correlating algorithms is often the state estimator \cite{Be:99,Ab:10}.

Two dynamics models will be presented in this paper, both with the actual parameters used in the numerical illustration section, where ePMFs and PF estimators are compared.

\subsection{General Model Definition}
The continuous dynamics (CD) is given by the state stochastic differential equation
	\begin{align}
	d\bfx(t)&=\bfA(t) \bfx(t) dt+  \bfu(t)dt +\bfQ_c(t) d\bfw(t)\label{eq:dynam},
	\end{align}
	where $\bfx(t)\in\real^{n_x}$ is the \textit{unknown} state of the system at (continuous) time $t$, $\bfu(t)$ is \textit{known} input, $\bfA \in \real^{n_x \times n_x}$ is \textit{known}  matrix, $\bfw(t)$ is the state noise modeled by the Brownian motion with the \textit{normally} distributed increment with the covariance matrix $E[d\bfw(t)(d\bfw(t))^T]=\bfI_{n_x}dt$, and $\bfQ_c(t)\in\real^{n_x\times n_x}$ is the matrix with \textit{known} diffusion coefficients. The state noise is supposed to be independent of the initial state $\bfx(0)$ with the  \textit{known} PDF $p(\bfx(0))$. 
	
	The discrete dynamics (DD) is given by the state stochastic difference equation
	\begin{align}
	\bfx_{t_{k+1}}&=\bfF_{t_k} \bfx_{t_k} + \bfu_k+ \bfw_{t_k},\label{eq:asx} 
	\end{align}
 where the meaning of the quantities is analogous to the CD case, $\bfF \in \real^{n_x \times n_x}$ is \textit{known} matrix, and $\bfw_{t_k}$ is the state noise with the known PDF $p(\bfw_{t_k})$.  For simplicity, the standard notation of discrete quantities, $ \mathbf{x}_{t_k} = \mathbf{x}_{k}$, is used. It is also assumed that $t_{k+1} - t_k = T_s = 1$. 
 
 The measurement equation is for both cases discrete
 	\begin{align}
	\bfz_{k}&=\bfh_{k}(\bfx_{k})+\bfv_{k},\label{eq:asz}
	\end{align}  
	$\bfz_{k}\in\real^{n_z}$ is the \textit{known} measurement at (discrete) time step $k$, $\bfh_k:\real^{n_x}\rightarrow\real^{n_z}$ is \textit{known} transformations, and $\bfv_{k}$ is the measurement noise with the known PDF $p(\bfv_{k})$.

	\subsection{Dynamic Random Walk Model}
	The first model used for comparison is the random walk model, the simplest model, with a two-dimensional state $\bfx_k = [ p_x \ p_y ]$ consisting of horizontal position coordinates of the vehicle in meters. Essentially it is a model with random walk dynamics and known velocity input $\bfu$ $[m/s]$ \cite{Be:99}. It can be used when the estimator design requires very simple model dynamics. The discrete dynamics parameters of the model are
	\begin{align}
		\bfF = \begin{bmatrix}
			1 & 0 \\ 0 & 1
		\end{bmatrix}, \bfu = \begin{bmatrix}
			50 \\ 50
		\end{bmatrix},\\
		p(\bfx_0) \sim \mathcal{N}\left\{\bfx_0;\begin{bmatrix}
			36569 \\
			55581
		\end{bmatrix}, \begin{bmatrix}
			160 & 20 \\
			20 & 90
		\end{bmatrix} \right\}, \\
		p(\bfw_k) \sim \mathcal{N}\{ \mathbf{w}_k ;\mathbf{0}, \bfQ_d \}, \bfQ_d = \diag(100),
	\end{align}
	where $\mathcal{N}\left\{\bfw_k; \bf0, \diag(100) \right\}$ is a Gaussian pdf of $\bfw_k$ with mean $\mathbf{0}$, and diagonal covariance matrix with $100$ on diagonal. 
	The equivalent continuous dynamics model therefore has
	\begin{align}
		\bfA = \mathbf{0}, \bfQ_c = \bfQ_d,
	\end{align} 
	and the same input and initial condition.
	

	\subsection{Dynamic Coordinated Turn with Known Rate}
	The second, more complex model, used for comparison has a four-dimensional state $\bfx_k = [p_x \ v_x \ p_y  \ v_y]$, which describes the horizontal position ($p_x, p_y$) $[m]$ and velocity ($v_x, v_y$) $[m/s]$ of the vehicle \cite{RoJi:03}. The parameters of the DD model are
		\begin{align}
		\bfF &=  \begin{bmatrix}
			1 & \frac{\sin(\alpha T_s)}{\alpha} & 0  & \frac{\cos(\alpha T_s) - 1}{\alpha} \\
			0 & \cos(\alpha T_s) & 0 &-\sin(\alpha T_s) \\
			0 & \frac{1-\cos(\alpha T_s)}{\alpha} & 1 & \frac{\sin(\alpha T_s)}{\alpha}\\
			0 & \sin(\alpha T_s) & 0 & \cos(\alpha T_s)
		\end{bmatrix}, \bfu = \mathbf{0}\\
		p(\bfx_0) &\sim \mathcal{N}\left\{\bfx_0;\begin{bmatrix}
			36569 \\
			50 \\
			55581\\
			50
		\end{bmatrix}, \begin{bmatrix}
			90 & 0 & 0 & 0 \\
			0 & 160 & 0 & 0 \\
			0 & 0 & 5 & 0 \\
			0 & 0 & 0 & 5 
		\end{bmatrix} \right\}, \\
		p(\bfw_k) &\sim \mathcal{N}\{ \mathbf{w}_k ;\mathbf{0}, \bfQ_d \}, \\\bfQ_d
 &=
\left[\begin{matrix}
 \frac{2(\alpha T_s - \sin(\alpha T_s))}{\alpha^3} & \frac{1-\cos(\alpha T_s)}{\alpha^2}\\
 \frac{1- \cos(\alpha T_s)}{\alpha^2} & T\\
  0 & \frac{-\alpha T_s - \sin(\alpha T_s)}{\alpha^2}\\
  \frac{\alpha T_s - \sin(\alpha T_s)}{\alpha^2} & 0
  \end{matrix}\right.\nonumber\\
&\qquad\qquad
\left.\begin{matrix}
  0 & \frac{\alpha T_s - \sin(\alpha T_s)}{\alpha^2}\\
  -\frac{\alpha T_s - \sin(\alpha T_s)}{\alpha^2} & 0\\
  \frac{2(\alpha T_s - \sin(\alpha T_s))}{\alpha^3} & \frac{1- \cos(\alpha T_s)}{\alpha^2}\\
  \frac{1-\cos(\alpha T_s)}{\alpha^2} & T_s
\end{matrix}\right], \label{eq:asx2}
	\end{align}
	where, $T_s = 1$ is the time step, and $\alpha = 30^{\circ}$ is the known turn rate.

	
	\subsection{TAN Measurement Equation}
	 The measurement equation is the same for both models. The measurement function $h$ is a discrete terrain map\footnote{The map is from Shuttle Radar Topography Mission (SRTM) an international project spearheaded by the U.S. National Geospatial-Intelligence Agency (NGA) and the U.S. National Aeronautics and Space Administration (NASA), see https://www2.jpl.nasa.gov/srtm/index.html.} represented by a table function that assigns vertical position (i.e. altitude) to each combination of latitude and longitude it covers. The measurement $z_k$ is a terrain altitude below the vehicle which can be based on a barometric altimeter. 
	 
	 The noise $v_k$ is distributed according to Gaussian mixture PDF with two components (this is simulating terrain with unmapped bridge or tunnel)\cite{NoGu:09}
	\begin{align}
	p(v_k)=\sum_{g=1}^{2}\frac{1}{2} \mathcal{N}\{v_k;\hat{v}_g,P_g\},\label{eq:ex1_pdf}
	\end{align}
 where the particular means and covariance matrices are given as follows
$\hat{v}_1 = 0, \hat{v}_2 = 20,$ $P_1 = P_2 = 1.$

	\subsection{State Estimation}
	The goal of the state estimation in the Bayesian framework is to find the filtering PDF of the state $\bfx_k$ conditioned on all measurements $\bfz^k=[\bfz_0,\bfz_1,\ldots,\bfz_k]$\footnote{Even thought mentioned models have a scalar measurement, the theory is given for general $n_z$ dimensions.}  up to the time instant $k$, i.e., the conditional PDF $p(\bfx_k|\bfz^k), \forall k$, is sought.

The general solution to the \textit{discrete} state estimation is given by the BRRs for the conditional PDFs computation \cite{AnMo:79}
\begin{align}
p(\bfx_k|\bfz^k)&=\frac{p(\bfx_k|\bfz^{k-1})p(\bfz_k|\bfx_k)}{p(\bfz_k|\bfz^{k-1})},\label{eq:filt}\\
p(\bfx_{k}|\bfz^{k-1})&=\int p(\bfx_{k}|\bfx_{k-1})p(\bfx_{k-1}|\bfz^{k-1})d\bfx_{k-1},\label{eq:pred}
\end{align}
where $p(\bfx_{k}|\bfz^{k-1})$ is the one-step predictive PDF computed by the Chapman-Kolmogorov equation (CKE) \eqref{eq:pred} and $p(\bfx_k|\bfz^k)$ is the filtering PDF computed by the Bayes' rule \eqref{eq:filt}. The PDFs $p(\bfx_{k}|\bfx_{k-1})$ and $p(\bfz_{k}|\bfx_{k})$ are the state transition PDFs obtained from \eqref{eq:asx}, resp. \eqref{eq:asx2} and the measurement PDF obtained from \eqref{eq:asz}. The PDF $p(\bfz_k|\bfz^{k-1})=\int p(\bfx_k|\bfz^{k-1})p(\bfz_k|\bfx_k)$ $d\bfx_k$ is the one-step predictive PDF of the measurement. The estimate of the state is given by filtering and the predictive PDFs. The recursion \eqref{eq:filt}, \eqref{eq:pred} starts from $p(\bfx_0|\bfz^{-1})=p(\bfx_0)$.

In the case of the continuous dynamics model, the time update step is given by the Fokker-Planck equation (FPE)\footnote{The $p_{\mathbfcal{X}(t)}(\bfx)$ may or may not be conditioned by measurement $\bfz$. The random variable notation $\mathbfcal{X}(t)$ is stressed here to make the partial derivative clear, i.e. $p_{\mathbfcal{X}(t)}(\bfx)$ is an equivalent notation to $p(\bfx(t))$.}
\begin{align}
 	{\pderiv{p_{\mathbfcal{X}(t)}(\bfx)}{t}} &= -{\nabla\cdot \left(\bfA \bfx\ p_{\mathbfcal{X}(t)}(\bfx) \right)}\nonumber\\
 	& + \frac{1}{2}\nabla\cdot\left(\bfQ_c\left(\nabla^T p_{\mathbfcal{X}(t)}(\bfx)\right) \right), \label{eq:fokker}
\end{align}
where $t\in(k,k+1)$, $\nabla$ denotes the gradient operator as a row vector, and $\nabla \cdot$ is the divergence. In \eqref{eq:fokker}, the first right-hand side term is named \textit{hyperbolic} and it describes the \textit{advection} of the PDF tied to the state dynamics. The second term is named \textit{parabolic} and it describes the \textit{diffusion} caused by the state noise. Compared to the CKE \eqref{eq:pred}, the FPE \eqref{eq:fokker} holds for the Gaussian state noise only. For a non-Gaussian noise, the FPE would have an infinite number of terms on the right-hand side \cite{DuSpa:05}.

	\section{Point-Mass Filter}
The PMF is based on an approximation of a conditional PDF $p(\bfx_k|\bfz^m)$ (resp. $p(\bfx(t)|\bfz^m)$) by a \textit{piece-wise constant} point-mass density $\hat{p}(\bfx_{k}|\bfz^m;\bfxi_k)$ defined at the set of the discrete grid points\footnote{MATLAB\textregistered\ style notation is used throughout the paper for simple comparison with published codes, where  $j$-th element of vector $\bfx_k$ is denoted as $\bfx_k^{(j)}$ and element of matrix $\bfxi_k$ at $i$-th row and $j$-th column is denoted as $\bfxi^{(j,i)}_k$.} $\bfxi_k=[\bfxi^{(:,1)},...,\bfxi^{(:,N)}_k], \bfxi^{(:,i)}_k\in\real^{n_x}$, as follows 
\begin{align}
\hat{p}(\bfx_k|\bfz^m;\bfxi_k)\triangleq\sum_{i=1}^NP_{k|m}(\bfxi^{(:,i)}_k)S\{\bfx_k;\bfxi^{(:,i)}_k,\bfDelta_k\},\label{eq:pdf_pm}
\end{align}
with
\begin{itemize}[leftmargin=\parindent,align=left,labelwidth=\parindent,labelsep=0pt,noitemsep]
	\item $P_{k|m}(\bfxi^{(:,i)}_k)=c_k\tilde{P}_{k|m}(\bfxi^{(:,i)}_k)$, where $\tilde{P}_{k|m}(\bfxi^{(:,i)}_k)=p(\bfxi^{(:,i)}_k|\bfz^m)$ is the value of the conditional PDF $p(\bfx_k|\bfz^m)$ evaluated at the $i$-th grid point (also called PMD point weight) $\bfxi^{(:,i)}_k$, $c_k=\delta_k\sum_{i=1}^{N}\tilde{P}_{k|m}(\bfxi^{(:,i)}_k)$ is a normalisation constant, and $\delta_k$ is the volume of the $i$-th point neighbourhood defined below,
	\item $\bfDelta_k$ defines a (hyper-) rectangular neighbourhood of a grid point $\bfxi^{(:,i)}_k$, where the PDF $p(\bfx_k|\bfz^m)$ is assumed to be constant and has value $P_{k|m}(\bfxi^{(:,i)}_k)$, and
	\item $S\{\bfx_k;\bfxi^{(:,i)}_k,\bfDelta_k\}$ is the \textit{selection} function defined as
	\begin{align}
	S\{\!\bfx_k;\bfxi^{(:,i)}_k\!,\!\bfDelta_k\!\}\!=\!\begin{cases}
	\!1,\mathrm{if}\ |\bfx_k^{(j)}\!-\!\bfxi^{(j,i)}_k|\!\leq\!\tfrac{\bfDelta_k^{(j)}}{2}\forall j,\\
	\!0, \mathrm{otherwise}.
	\end{cases}\!\!\!\!\label{eq:sf3}
	\end{align}
	so that 
	$\int S\{\bfx_k;\bfxi^{(:,i)}_k,\bfDelta_{k}\}d\bfx_k=\prod_{i=1}^{n_x}\bfDelta_k^{(i)}=\delta_k$. 
\end{itemize}


The basic algorithm of the PMF is summarised as \cite{SiKraSo:06}:
\newline
\rule{8.5cm}{0.5pt}
\newline
\textbf{Algorithm 1: Point-Mass Filter}
\vspace*{-1mm}
\begin{enumerate}[leftmargin=\parindent,align=left,labelwidth=\parindent,labelsep=0pt,noitemsep]
	\item \textit{Initialisation}: Set $k=0$, construct the initial grid of points $\{\bfxi_0^{(:,i)}\}_{i=0}^N$, and define the initial point-mass PDF $\hat{p}(\bfx_0|\bfz^{-1};\bfxi_0)$ of form \eqref{eq:pdf_pm} 
	approximating the initial PDF.
	\item  \textit{Meas. update}: Compute the filtering PMD $\hat{p}(\bfx_k|\bfz^{k};\bfxi_k)$ of the form \eqref{eq:pdf_pm} 
	where the PDF value at $i$-th grid point is
	$P_{k|k}(\bfxi^{(:,i)}_k)=\tfrac{p(\bfz_k|\bfx_k=\bfxi^{(:,i)}_k)P_{k|k-1}(\bfxi^{(:,i)}_k)}{\sum_{i=1}^Np(\bfz_k|\bfx_k=\bfxi^{(:,i)}_k)P_{k|k-1}(\bfxi^{(:,i)}_k)\delta_k}$.
	\item \textit{Grid construction:} Construct the new\footnote{The number of grid points $N$ is kept constant $\forall k$ to ensure constant (and predictable) computational complexity of the PMF.} grid $\{\bfxi_{k+1}^{(j)}\}_{j=0}^N$.
	\item \textit{Time update}: Compute the predictive point-mass PDF $\hat{p}(\bfx_{k+1}|\bfz^{k};\bfxi_{k+1})$ of the form \eqref{eq:pdf_pm} at the new grid of points 
	where the weight of the predictive PMD at $j$-th grid point is
	\begin{itemize}
		\item solution of numerical method \cite{MaDuBrEl:21,MaBrDu:21,KaSch13} in the continuous case, and
		\item $P_{k+1|k}(\bfxi^{(:,j)}_{k+1})= \overset{N}{\underset{i=1}{\sum}} p(\bfxi^{(:,j)}_{k+1}|\bfx_k=\bfxi^{(:,i)}_{k})P_{k|k}(\bfxi^{(:,i)}_k)\delta_k$ in the discrete case.
	\end{itemize}
	\item Set $k=k+1$ and go to step 2).
\end{enumerate}
\vspace*{-3mm}
\rule{8.5cm}{0.5pt}
\newline
\vspace*{-3mm}

%

From step 4, it can be seen, that the time evolution of the PMD can be written in a matrix form as a product of transition probability matrix $\bfT_k$ (TPM) and filtering PMD ($P_{{k}}^{(:)}$), that is
	\begin{align}
	P_{{k+1}}^{(:)} =\bfT_k\ P_{{k}}^{(:)}, \label{eq:TPM}
	\end{align}
where notation $P_{{k}}^{(:)}$ stands for all weights $P_{{k}}^{(i)}\ \forall i$ stacked in a column vector, $\bfT_k$ is a TPM, which has values equal to $p(\bfxi^{(:,j)}_{k+1}|\bfx_k=\bfxi^{(:,i)}_{k})$, in the discrete case, and given by numerical method in the continuous case (see further).




	\subsection{Computational Complexity Reduction}
	A range of techniques for the PMF computational complexity reduction was proposed, however, at the cost of additional approximations, the need for user/designer defined parameters, or for models of a special form. To name a few: \textit{Rao-Blackwellisation} \cite{SmGa:13,DuSoVeStHa:19}, \textit{Separable prediction} \cite{Be:99}, \textit{Copula prediction} \cite{DuStMaBl:22}, \textit{Tensor-based prediction} \cite{TiStDu:22,LiWaYaZh:19}. Also, note that the PMF grid can be designed in an adaptive or sparse layout. Although those layouts lead to the reduction of the total number of grid points $N$, the order of the computational complexity is still $\mathcal{O}\left( N^2 \right)$ \cite{KaSch13}.

		\section{Efficient Point-Mass Prediction}
		The key idea in the ePMF is to transform the time-update from a general product of a matrix and a vector \eqref{eq:TPM} to a convolution, therefore allowing the usage of computationally efficient convolution theorem. Using the CD estimation terminology, it can be said that the solution to a diffusion part of the estimation is (for an equidistantly spaced grid) a convolution, but the solution to the advection part (in both DD and CD case) is not, for a general grid movement, a convolution. Therefore, the idea is to solve the advection by particular grid movement and to solve the diffusion by a PMF update (CD or DD). The grid movement, not only solves the advection problem but also ensures the proper grid placement with respect to state dynamics.
		
		For an arbitrary grid movement, the time update cannot be written as a convolution. That is why the ePMF uses a specific grid movement \cite{MaDuBrPaCh:23}
	\begin{align}
	\bfxi_{k+1}^{(:,i)} &= \bfF\bfxi_{k}^{(:,i)}, \forall i\label{eq:newGridDD}
\end{align}
for the DD model, and 
\begin{align}
	 \bfxi_{t_{k+1}}^{(i,:)} &= \exp(\bfA T_s) \bfxi_{t_k}^{(:,i)}, \forall i\label{eq:forcMovGrid}
\end{align}
for the CD model.

\subsection{Fast Fourier Transform Based Solution}

\begin{figure}
		\includegraphics[width=0.5\textwidth]{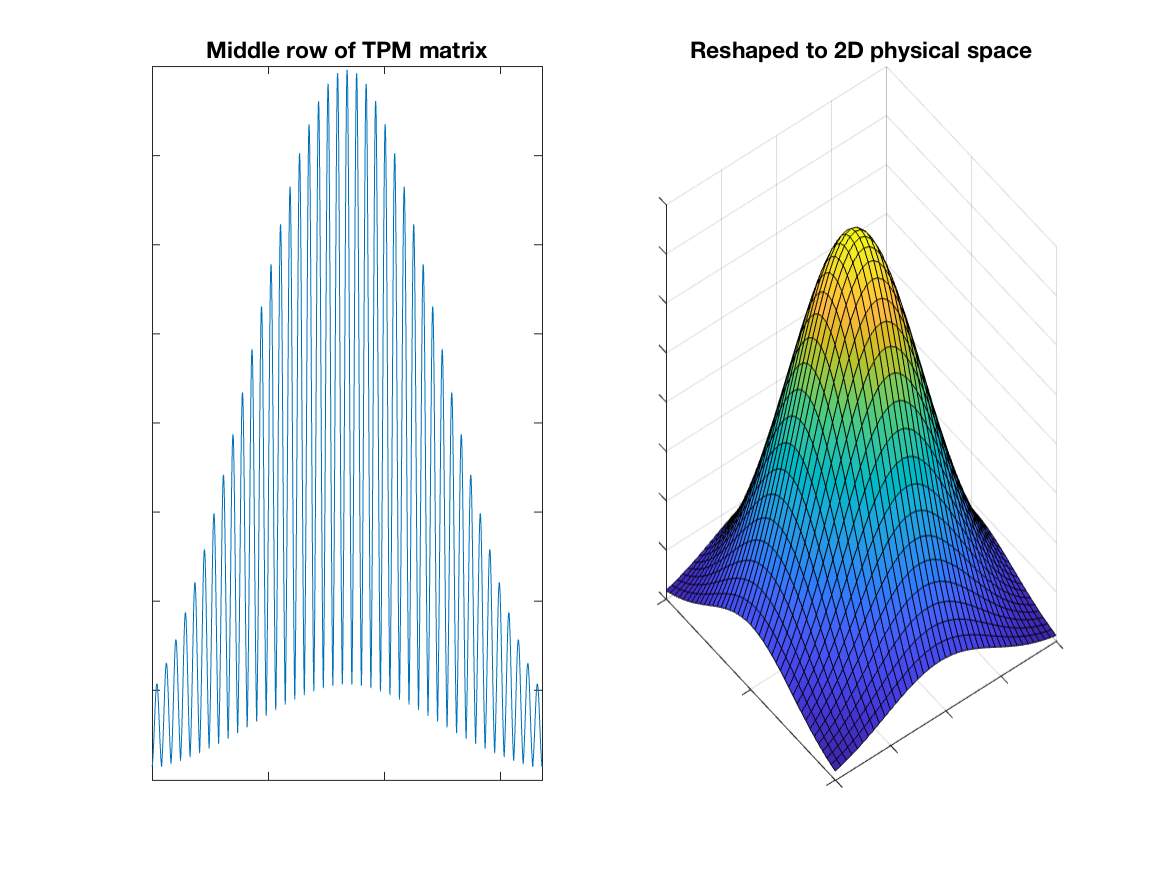}
		\caption{Computational to physical space re-shaping for 2D space and Gaussian noise}
		\label{fig:reshape}
	\end{figure}

For the grid movement \eqref{eq:newGridDD}, \eqref{eq:forcMovGrid}, the TPM matrix rows contain the same unique values of the Gaussian transition kernel with mean $\bfxi_{k+1}^{(:,j)}$ for $j$-th row, therefore the predictive weights can be calculated by $n_x$ dimensional convolution with zero padding as
\begin{align}
	\widetilde{P}_{{k+1}}^{(:)} = \widetilde{\bfT^{(m,:)}_k} * \overset{n_x}{...} *  \widetilde{P}_{{k}}^{(:)}. \label{eq:tpmconv}
\end{align}
where the symbol $* \overset{n_x}{...} *$ denotes the convolution in $n_x$-dimensional space, and $\widetilde{\bfT^{(m,:)}_k}$ denotes the middle row of the TPM ($m = \left\lceil \frac{N}{2} \right\rceil $). Matrices with $\sim$ overhead are transformed from computational space generally to tensors in the physical space as shown in Figure \ref{fig:reshape} for 2 dimensions with Gaussian noise (i.e. for model \eqref{eq:asx}). 

The convolution theorem can be now applied on \eqref{eq:tpmconv}, for efficient calculation in the frequency domain
	\begin{align}
	\widetilde P_{{k+1}}^{(:)} = \mathcal{F}^{-1}\left(  \mathcal{F}(\widetilde{\bfT^{(m,:)}_k}) \odot \mathcal{F}(\widetilde{ P}_{{k}}^{(:)})\right), \label{eq:tpmconvFFT}
	\end{align}
	where $\mathcal{F}$ denotes the Fourier transform and $\odot$ the Hadmard product. Let's call this method FFT-PMF.
	
	For the DD model, the values of the TPM middle row can be directly found by evaluating the transition PMD. However, for the CD system, the procedure is a bit more complicated, and it is illustrated for $n_x = 1$.

Using the explicit central difference method to solve the remaining diffusion part of the FPE, the finite difference operator is a tri-diagonal matrix
	\begin{align}
	\bfD_{{t_k}} = \begin{bmatrix}
	b  & a &  &   \\
	a & b  & a & & \  \\
	& a & b & a &    & \\
	&  & \ddots & \ddots &  \ddots   & \\
	\end{bmatrix},
	\end{align}
	where $a = \frac{Q_c\ \Delta t}{2\Delta_{t_k} ^2}$, and $b = 1 - \frac{Q_c \Delta t}{\Delta_{t_k} ^2} - \Delta t  A$. The $\Delta t  A$ is a constant part of the advection that was not solved by the grid movement, and $\Delta t$ is the numerical method time step. Thus one numerical time step can be performed as 
	\begin{align}
		P_{{t_k+\Delta t}}^{(:)} = \bfD_{{t_k}} P_{{t_k}}^{(:)}.
	\end{align}
	Then the whole update from time $t_k$ to time $t_{k+1}$ is
	\begin{align}
	P_{{t_{k+1}}}^{(:)} =\underbrace{\bfD_{t_k+(l-1)\Delta_t} \cdots \bfD_{t_k+\Delta_t}\bfD_{t_k}}_{\bfT_k}\ P_{{t_k}}^{(:)}. \label{eq:FKEnumSol}
	\end{align}
 Therefore to find the TPM middle row values one would have to calculate the 'power' of the matrix $\bfD_{t_k}$ (leading to $\bfT_k$), which would be nearly impossible for a higher number of grid points. Therefore an efficient calculation of only the middle row of TPM $\bfT_k$ in the continuous case is proposed.

	\subsection{Fast Sine Transform Based Solution}
	
	The matrix $\bfT_k$ can be efficiently calculated using the eigenvalue and eigenvector form as
	\begin{align}
	\bfT_k = (\bfR \bfLambda_{t_k}\bfR^{-1})^l,\label{eq:decomp}
	\end{align}
	where $\bfR$ is a matrix of eigenvectors as columns, and $\Lambda_{t_k}$ is a diagonal matrix with eigenvalues $\lambda_{t_k}$ on diagonal.
		Because the matrix $\bfD_{t_k}$ (resp. $\bfT_k$) is tri-diagonal\footnote{Can be extended for general Toeplitz matrices arising in multidimensional estimation \cite{MaDuBrPaCh:23}.}, its $j$-th eigenvalue is \cite{Sa:06}
	\begin{align}
	\lambda^{(j)}_{t_k} = b + 2a \cos\left(\frac{j\pi}{N+1} \right),
	\end{align}
	and $j$-th eigenvector
	\begin{align}
	\bfR^{(:,j)} = \begin{bmatrix}
	\sin(\frac{1j\pi}{N+1})\\
	\vdots \\
	\sin(\frac{Nj\pi}{N+1})\\
	\end{bmatrix},\ j=1,...,N.
	\end{align}
	For the eigenvector matrix, it also holds that 
	\begin{align}
	\bfR^{-1} = \frac{2}{N+1} \bfR.
	\end{align}	
	As the eigenvalues $\lambda_{t_k}$ are time-varying because of the grid movement \eqref{eq:newGridDD}, \eqref{eq:forcMovGrid}, and the eigenvectors are time-invariant (i.e., constant), the decomposition \eqref{eq:decomp} can be treated as	
	\begin{align}
	\bfT_k = \bfR \underbrace{\left(\bfLambda_{t_k+(l-1)\Delta t}\odot \cdots \odot\bfLambda_{t_k+\Delta t} \odot \bfLambda_{t_k} \right)}_{\bfLambda_{\text{pow}}}\bfR^{-1}.
	\end{align}

	Therefore, calculation of an arbitrary element $\bfT^{(j,i)}_k$ reads
	\begin{align}
	\bfT^{(i,j)}_k = \frac{2}{N+1} \sum_{s=1}^N \bfLambda_{\text{pow}}^{(s,s)} \sin\left( \frac{is\pi}{N+1} \right) \sin\left( \frac{js\pi}{N+1} \right). \label{dzij}
	\end{align}

	The extension to $n_x$ dimensions can be found in our previous paper \cite{MaDuBrPaCh:23}.
	The proposed solution is a combination of $n_x$ one-dimensional solutions, based on the Kronecker product. 
	For two dimensions and diagonal state noise covariance matrix $\bfQ_c$, the finite difference matrix is block tridiagonal
	\begin{align}
	{\bfD}_{t_k} = \begin{bmatrix}
	\bfK & \bfC &   &   \\
	\bfC & \bfK  & \bfC &  &  \  \\
	& \bfC & \bfK &  \bfC&    &  \\
	&  & \ddots & \ddots &  \ddots   & \\
	\end{bmatrix} \in \mathbb{R}^{N \times N}, \label{eq:2Ddiff}
	\end{align} 
	\begin{align}
	\bfK = \begin{bmatrix}
	b  & a &  &   \\
	a & b  & a & & \  \\
	& a & b & a &    & \\
	&  & \ddots & \ddots &  \ddots   & \\
	\end{bmatrix}\in \mathbb{R}^{N \times N},
	\end{align}
	where $a = \frac{\bfQ_c^{(1,1)}\ \Delta t}{2\Delta_{t_k}^2(1) }$, $b = 1 - \frac{\bfQ_c^{(1,1)}\Delta t}{\Delta_{t_k}^2(1)}  - \frac{\bfQ_c^{(2,2)}\Delta t}{\Delta_{t_k}^2(2)}- \Delta t \ \operatorname{trace}(\bfA)$, and $\bfC$ is a diagonal matrix with elements  $c=\frac{\bfQ_c^{(2,2)}\ \Delta t}{2\Delta_{t_k}^2(2) }$.
	
	The eigenvalues of \eqref{eq:2Ddiff} can then be calculated as 
	\begin{align}
	b + 2\ a \cos \left( \frac{i\pi}{N_1+1} \right) +  2\ c \cos \left( \frac{j\pi}{N_2+1} \right),\\
	i = 1,2,...,N_1\nonumber \\ \nonumber
	j = 1,2,...,N_2,
	\end{align}
	where $N^{(1)}$ is the number of points on first dimension axis ($N = \prod_{i=1}^{n_x} N^{(i)}$). The diagonal matrix $\bfLambda_{t_k}$ is formed from the eigenvalues. The eigenvectors can be extended to $n_x$  in a similar way \cite{MaDuBrPaCh:23}.
%

	The equation \eqref{dzij} (and analogous expression for $n_x$) allows the computation of an arbitrary element of the TPM as easily as in the DD case, therefore the convolution theorem can now be used \eqref{eq:tpmconvFFT}. However, on a closer inspection, it can be seen that calculation of ${\bfT}^{(m,:)}_k$ \eqref{dzij} can be rewritten to two discrete sine transforms \cite{St:07}, therefore in the CD case the time update can be directly calculated as
	\begin{align}
	\widetilde P_{{k+1}}^{(:)} = \mathcal{S} \left( \widetilde{\diag(\bfLambda_{\text{pow}})} \odot \mathcal{S}(\widetilde{ P}_{{k}}^{(:)}) \right),
\end{align}
where $\mathcal{S}$ denotes the fast sine transform, and $\widetilde{\diag(\bfLambda_{\text{pow}})}$ is a tensor (in general $n_x$) formed by taking the diagonal values of $\bfLambda_{\text{pow}}$ and reshaping them to physical space. Let's call this method FST-PMF. Note that the FST-PMF is using one less (inverse) fast transform, and thus can be less prone to numerical errors, and less computationally complex. The extension to higher dimensions is following the same pattern.

	
	\section{Efficient Point-Mass Filter Design}
In order for efficient prediction to be used during the state estimation routine, some issues have to be addressed. 
	
	\subsection{Grid Movement Compensation}
	The grid movement is based solely on the state dynamics, therefore the influence of the state noise is ignored. Also, the grid movement is similar to the particle filter time update, therefore there arises a similar degeneracy problem as in the particle filter without resampling \cite[p. 40]{RiAr:04}. That is why the grid generally has to be re-designed outside the time-update step. The grid is re-designed on the basis of sigma probabilities, just before the time-update grid movement. The PMD weights over the new grid are calculated by interpolation.
	

\noindent\rule{8.5cm}{0.5pt}
\newline
\textbf{Algorithm 2: Grid Re-design}
\vspace*{-1mm}
\begin{enumerate}[leftmargin=\parindent,align=left,labelwidth=\parindent,labelsep=0pt,noitemsep]
\setcounter{enumi}{0}
	\item \textit{Approximate predictive moments}: The predictive first two moments are calculated by Kalman Filter $\hat{\bfx}_{k+1|k}$, and ${\bfP}_{k+1|k}$.
	\item \textit{Set corners of required predictive grid $\bfc_i^{\text{pred}}$:} The grid is based on the predictive moments and Chebyshev inequality/$\sigma$ ellipse probability. Therefore its middle is at $\hat{\bfx}_{k+1|k}$, and size is based on ${\bfP}_{k+1|k}$.
	\item \textit{Transform the corners to filtration density space:} $\bfc_i^{\text{meas}} = \bfF^{-1} \bfc_i^{\text{pred}}$.
	\item \textit{Design the new filtering grid $\bfxi_{k}^{\text{new}}$:} It is designed, so that it is as small as possible while circumscribing all corners $\bfc_i^{\text{meas}}$, and having boundaries aligned with state-space axes (reduces the interpolation computational complexity at the cost of excessive coverage of state-space).
	\item \textit{Interpolate the filtering density weights on $\bfxi_{k}^{\text{new}}$:} Interpolate the weights of $P_{k|k}(\bfxi^{\text{new},(:,i)}_k)$ from $P_{k|k}(\bfxi^{(:,i)}_k)$, respectively $\bfF^{-1}\left(P_{k|k}(\bfxi^{\text{new},(:,i)}_k)\right)$ from $\bfF^{-1}\left(P_{k|k}(\bfxi^{(:,i)}_k)\right)$ (see next paragraph). 
\end{enumerate}
\vspace*{-3mm}
\rule{8.5cm}{0.5pt}
\newline
\vspace*{-3mm}

\begin{figure*}
		\centering\includegraphics[width=0.68\textwidth]{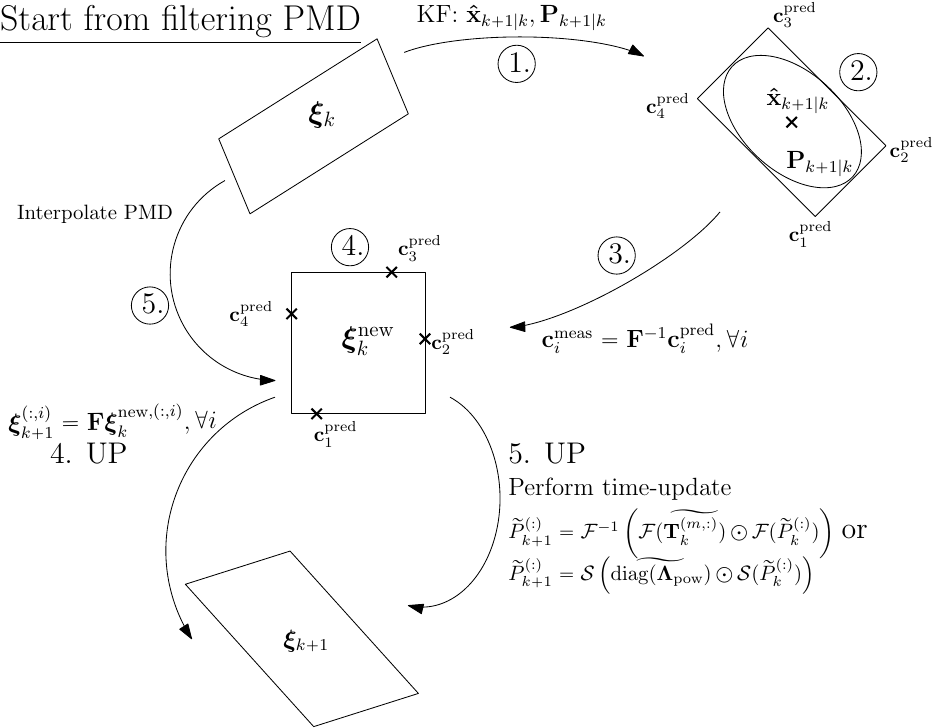}
		\caption{Scheme of the grid re-design}
		\label{fig:grids}
	\end{figure*}

For the CD case a dynamics $ \exp(\bfA T_s)$ is used instead of $\bfF$. The grid re-design is depicted in Figure \ref{fig:grids}, where the steps in circles are corresponding to the Algorithm 2 steps, and steps with 'UP' correspond to the Algorithm 3 steps.

In the grid re-design step an interpolation of filtering density is needed. The filtering PMD has to be interpolated from the filtering grid $P_{k|k}(\bfxi^{(:,i)}_k)$ to the re-designed grid $P_{k|k}(\bfxi^{\text{new},(:,i)}_k)$. Unfortunately, the filtering grid (i.e. predictive grid from last step $k-1$ to this step $k$) has generally a rhomboid size, because of \eqref{eq:newGridDD}, and \eqref{eq:forcMovGrid}. This means that a computationally expensive MATLAB\textregistered\ function \textit{scatteredInterpolant} has to be used.

 Fortunately, an alternative approach is proposed to significantly reduce computational complexity. Instead of interpolating $P_{k|k}(\bfxi^{\text{new},(:,i)}_k)$ from $P_{k|k}(\bfxi^{(:,i)}_k)$ a $\bfF^{-1}\left(P_{k|k}(\bfxi^{\text{new},(:,i)}_k)\right)$ can be interpolated from $\bfF^{-1}\left(P_{k|k}(\bfxi^{(:,i)}_k)\right)$ leading to a same result. Now $\bfF^{-1}\left(P_{k|k}(\bfxi^{(:,i)}_k)\right)$ is a rectangular grid with boundaries aligned with state axes. That allows the usage of computationally efficient MATLAB\textregistered\ function called $griddedInterpolant$, which is working only on structured grids.


	\subsection{Design of Continuous ePMF for non-Diagonal Noise}

	In the case of a non-diagonal noise, the diffusion matrix does not have the special form needed for the eigenvalues and eigenvectors to be easily calculated as mentioned. Fortunately, a noise diagonalization can be performed. Suppose a model of a form \eqref{eq:dynam}
	where $\bfQ_c$ is a full matrix (i.e. with nonzero non-diagonal elements). Instead of solving the FPE for this model, an alternative FPE can be solved
	\begin{align}
 	{\pderiv{p_{\mathbfcal{X}(t)}(\bar{\bfx})}{t}} &= -{\nabla_{\bar{\bfx}}\cdot \left(\bar{\bfA} \bar{\bfx}\ p_{\mathbfcal{X}(t)}(\bar{\bfx}) \right)}\nonumber\\
 	& + \frac{1}{2}\nabla_{\bar{\bfx}}\cdot\left(\nabla_{\bar{\bfx}}^T p_{\mathbfcal{X}(t)}(\bar{\bfx})\right), \label{eq:fokker2}
\end{align}
where a substitution was done, $\bfQ = \bfG \bfG^T$, $\bar{\bfx}= \bfG^{-1} \bfx$, $\bar{\bfA} = \bfG^{-1}\bfA\bfG$, and $\nabla_{\bar{\bfx}} = [ \frac{\partial}{\partial \bar{\bfx}(1)} ... \frac{\partial}{\partial \bar{\bfx}(n_x)}]$.

Then the estimation algorithm is run in the state space $\bar{\bfx}$. If the first two moments (i.e. the estimate and its uncertainty) for the original FPE solution are needed they can be calculated as $\mean[\bfx] = \bfG\mean[\bar{\bfx}]$, $\cov[\bfx] = \bfG\ \cov[\bar{\bfx}]\ \bfG^{-1}$.
	
	\subsection{Time Varying Systems}
Models with time-dependent dynamics $\bfF_k$, or noise characteristics $\bfQ_{k}$ do not create any issues for the ePMF algorithm, it just has to be kept in mind to change the grid movement $\bfxi_{k+1}^{(:,i)} = \bfF_k\bfxi_{k}^{(:,i)}, \forall i$, corresponding grid space step discretization $\Delta_k$, and calculate the transition probability values ${\bfT^{(m,:)}_k}$ using correct $\bfQ_{d,k}$. For the CD the case is analogous.

		\subsection{Algorithm}
	The final ePMF algorithm is as follows.	
\noindent\rule{8.5cm}{0.5pt}
\newline
\textbf{Algorithm 3: Efficient Point-Mass Filter}
\vspace*{-1mm}
\begin{enumerate}[leftmargin=\parindent,align=left,labelwidth=\parindent,labelsep=0pt,noitemsep]
\setcounter{enumi}{0}
	\item \textit{Initialisation}: Set $k=0$, construct the initial grid of points $\{\bfxi_0^{(:,i)}\}_{i=0}^N$, and define the initial point-mass PDF $\hat{p}(\bfx_0|\bfz^{-1};\bfxi_0)$ of form \eqref{eq:pdf_pm} 
	approximating the initial PDF.
	\item  \textit{Meas. update}: Compute the filtering PMD $\hat{p}(\bfx_k|\bfz^{k};\bfxi_k)$ of the form \eqref{eq:pdf_pm} 
	where the PDF weight at $i$-th grid point is
	$P_{k|k}(\bfxi^{(:,i)}_k)=\tfrac{p(\bfz_k|\bfx_k=\bfxi^{(:,i)}_k)P_{k|k-1}(\bfxi^{(:,i)}_k)}{\sum_{i=1}^Np(\bfz_k|\bfx_k=\bfxi^{(:,i)}_k)P_{k|k-1}(\bfxi^{(:,i)}_k)\delta_k}$.
	
	\item If $k>0$ \ \textit{Grid re-design:} See Algorithm 2. 

	\item \textit{Grid movement:} Construct the predictive\footnote{The number of grid points $N$ is kept constant $\forall k$ to ensure constant (and predictable) computational complexity of the PMF.} grid $\bfxi_{k+1}^{(:,j)} =  \bfF\ \bfxi_{k}^{\text{new},(:,j)}\ \forall j$, or in CD case $\bfxi_{t_{k+1}}^{(:,i)} = \exp(\bfA T_s) \bfxi_{t_k}^{(:,i)}$.
		\item \textit{Time update}: Compute the predictive weights $\widetilde P_{{k+1}}^{(:)} = \mathcal{F}^{-1}\left( \mathcal{F}(\widetilde{\bfT^{(m,:)}_k}) \odot \mathcal{F}(\widetilde{ P}_{{k}}^{(:)})\right)$, or in CD case alternatively as $	\widetilde P_{{k+1}}^{(:)} = \mathcal{S} \left( \widetilde{\diag(\bfLambda_{\text{pow}})} \odot \mathcal{S}(\widetilde{ P}_{{k}}^{(:)}) \right)$.
		\item Set $k=k+1$ and go to step 2).
\end{enumerate}
\vspace*{-3mm}
\rule{8.5cm}{0.5pt}
\newline
\vspace*{-3mm}

 \subsection{On the Compatibility with State of the Art Algorithms}
 
 Because considered models have linear dynamics, and only one row of the TPM has to be calculated, the ePMF is ideal to be combined with the reliable convolution algorithm, to achieve higher accuracy of estimates \cite{DuStMaBr:21,DuStMa:20}. 
 
 Even higher computational complexity reduction can be achieved by employing the thrifty convolution \cite{SiKraSo:06} and computing only part of the middle TPM row.
  
The proposed algorithms have time-update in form of a convolution, this opens up ways to use theoretical findings usually used in other fields to further speedup the update at the cost of accuracy, for example, a fast Gauss transform can be used \cite{YaDuGuDa:03}.

\section{Enclosed MATLAB\textregistered\ Codes}
The codes are published on the website of our research team IDM (https://idm.kky.zcu.cz/sw.html).

Two main files are supplied \textit{main2D.m} and \textit{main4D.m}, these are containing the appropriate implementations for models from Section \ref{sec:tan}. The supplemented code uses the following publicly available mex files, these should be compiled first by running file \textit{runFirst\_\_\_.m}. The PF bootstrap implementation is using binary search published in \cite{Av:22}. The FFT-PMF implementation is using FFT-based convolution from \cite{Lu:22}. The FST-PMF is using the MATLAB\textregistered\ Discrete Trigonometric Transform Library from \cite{Tr:22}. The PMD grid creation routine is using the \textit{combvec.m} file available in \cite{Br:97}.

	\section{Numerical results}	
	The filters were used to estimate the state of the models presented in section \ref{sec:tan}, and compared using three criterions
\begin{itemize}
	\item Root-mean-square-error $\mathrm{RMSE}(j)=$
	\begin{align}
		\sqrt{\frac{1}{M(K+1)}\sum_{i=1}^M\sum_{k=0}^K((\bfx^{(j)}_k)^{[i]}-(\bfx^{(j)}_{k|k})^{[i]})^2},
	\end{align}
	\item Averaged standard deviation $\mathrm{aSTD}(j)=$
	\begin{align}
	\sqrt{\frac{1}{M(K+1)}\sum_{i=1}^M\sum_{k=0}^K{(\bfP^{(j,j)}_{k|k})^{[i]}}},
	\end{align}
	\item Average time in seconds of one filter step i.e. measurement and time update,
\end{itemize}
using $M=100$ Monte-Carlo (MC) simulations (same data were used for all estimators), where $(x^{(j)}_k)^{[i]}$ is the true state for $j$-th dimension at time $k$ at $i$-the MC simulation, $x_{k|k}^{[i]}=\mean[x_k^{[i]}|z^k]$ its filtering estimate, and $(P^{(j,j)}_{k|k})^{[i]}=\var[(x_k^{(j)})^{[i]}|z^k]$ the respective filtering variance diagonal element. Ideally, for an accurate estimate, the RMSE should be as small as possible and for a \textit{consistent} estimate, the RMSE and aSTD should be identical. The positive or negative difference between the RMSE and aSTD indicates \textit{optimistic} or \textit{pessimistic} estimate, respectively. 

The results for the random walk model can be seen in Table \ref{2D} for the number of grid points $N = 1681$, and the number of particles $N_p = 1681$. The results for the coordinated turn model can be seen in Table \ref{4D} for the number of grid points $N = 194,481$, and the number of particles $N_p = 1,200,000$. For illustration, in the 4D case, there is also included time of one step of a standard PMF. The numbers after FST-PMF are defining the numerical method time step size. The 4D results are only for the DD estimation as we do not possess mex files for calculating sine transform in 4D to effectively compare the CD solution to other solutions that either use MATLAB\textregistered\ native functions or mex files.

It can be seen that in two dimensions approximately $50\%$ of the computation time was saved. Moreover, in four dimensions approximately $99.9\%$ of time was saved. Also, the PMF is now comparable to PF in both accuracy and computational complexity, which was not possible before.

\begin{table*}[t]
\centering
\begin{tabular}{llllll}
& RMSE(1) & RMSE(2) & ASTD(1) & ASTD(2) & TIME \\ 
\hline 
PMF & 23.4517 & 16.2028 & 25.497 & 16.1807 & 0.11057 \\ 
eFFT & 23.4517 & 16.2028 & 25.497 & 16.1807 & 0.055116 \\ 
eSFT 0.01 & 23.3676 & 16.2094 & 24.7509 & 15.8515 & 0.055406 \\ 
eSFT 0.005 & 23.3971 & 16.2061 & 25.0858 & 15.9963 & 0.054584 \\ 
eSFT 0.001 & 23.4233 & 16.2047 & 25.3613 & 16.1161 & 0.055019 \\ 
PF bootstrap & 23.0559 & 16.9423 & 26.4885 & 18.2893 & 0.054651 \\ 
\hline 
\end{tabular}\caption{\label{2D}Two dimensional estimation results.}
\end{table*}
		
\begin{table*}[t]
\centering
			\begin{tabular}{llllllllll}
& RMSE(1) & RMSE(2) & RMSE(3) & RMSE(4) & aSTD(1) & aSTD(2) & aSTD(3) & aSTD(4) & TIME \\ 
\hline 
FFT-PMF &  25.777  & 18.339 & 18.751  & 12.065 &  28.436 & 21.634 & 20.741 & 13.555 & 0.35692\\ 
PF bootstrap & 34.375 & 28.571 & 21.801 & 16.11  & 47.878  & 43.583     & 28.124  & 22.567 & 0.38886 \\
PMF &-&-&-&-&-&-&-&-& 931.944 \\
\hline 
\end{tabular}\caption{\label{4D}Four dimensional estimation results.}
\end{table*}

	\section{Concluding Remarks}
	An efficient point-mass filter was proposed and illustrated on a terrain-aided navigation scenario for both CD and DD case. The resulting algorithm unifies CD and DD estimation, and it essentially combines the pros of PF and PMF as it is both deterministic/robust as PMF and fast as PF. The theoretical results were verified by numerical experiments. The paper also described enclose implementation files in MATLAB\textregistered .

	%
	%
	\bibliographystyle{IEEEtran}
	\bibliography{literatura}

\begin{thebibliography}{10}
\providecommand{\url}[1]{#1}
\csname url@samestyle\endcsname
\providecommand{\newblock}{\relax}
\providecommand{\bibinfo}[2]{#2}
\providecommand{\BIBentrySTDinterwordspacing}{\spaceskip=0pt\relax}
\providecommand{\BIBentryALTinterwordstretchfactor}{4}
\providecommand{\BIBentryALTinterwordspacing}{\spaceskip=\fontdimen2\font plus
\BIBentryALTinterwordstretchfactor\fontdimen3\font minus
  \fontdimen4\font\relax}
\providecommand{\BIBforeignlanguage}[2]{{%
\expandafter\ifx\csname l@#1\endcsname\relax
\typeout{** WARNING: IEEEtran.bst: No hyphenation pattern has been}%
\typeout{** loaded for the language `#1'. Using the pattern for}%
\typeout{** the default language instead.}%
\else
\language=\csname l@#1\endcsname
\fi
#2}}
\providecommand{\BIBdecl}{\relax}
\BIBdecl

\bibitem{So:74}
H.~W. Sorenson, ``On the development of practical nonlinear filters,''
  \emph{Information Sciences}, vol.~7, pp. 230--270, 1974.

\bibitem{AnMo:79}
B.~D.~O. Anderson and J.~B. Moore, \emph{Optimal Filtering}.\hskip 1em plus
  0.5em minus 0.4em\relax Prentice Hall, New Jersey, 1979.

\bibitem{JuUhl:04}
S.~J. Julier and J.~K. Uhlmann, ``Unscented filtering and nonlinear
  estimation,'' \emph{IEEE Proceedings}, vol.~92, no.~3, pp. 401--421, 2004.

\bibitem{DuStrSi:13}
J.~Dun\'{i}k, O.~Straka, and M.~\v{S}imandl, ``Stochastic integration filter,''
  \emph{IEEE Transactions on Automatic Control}, vol.~58, no.~6, pp.
  1561--1566, 2013.

\bibitem{JiXiChe:13}
B.~Jia, M.~Xin, and Y.~Cheng, ``High-degree cubature {K}alman filter,''
  \emph{Automatica}, vol.~49, no.~2, pp. 510--518, 2013.

\bibitem{SoAl:71}
H.~W. Sorenson and D.~L. Alspach, ``Recursive {B}ayesian estimation using
  {G}aussian sums,'' \emph{Automatica}, vol.~7, pp. 465--479, 1971.

\bibitem{DoFrGo-book:01}
A.~Doucet, N.~De~Freitas, and N.~Gordon, Eds., \emph{Sequential Monte Carlo
  Methods in Practice}.\hskip 1em plus 0.5em minus 0.4em\relax Springer, 2001,
  (Ed. Doucet A., de Freitas N., and Gordon N.).

\bibitem{Be:99}
N.~Bergman, ``Recursive bayesian estimation: Navigation and tracking
  applications,'' Ph.D. dissertation, Link\"{o}ping University, Sweden, 1999.

\bibitem{SiKraSo:02}
M.~\v{S}imandl, J.~Kr\'{a}lovec, and T.~S\"{o}derstr\"{o}m, ``Anticipative grid
  design in point-mass approach to nonlinear state estimation,'' \emph{IEEE
  Transactions on Automatic Control}, vol.~47, no.~4, 2002.

\bibitem{KaSch13}
C.~{Kalender} and A.~{Schottl}, ``Sparse grid-based nonlinear filtering,''
  \emph{IEEE Transactions on Aerospace and Electronic Systems}, vol.~49, no.~4,
  pp. 2386--2396, 2013.

\bibitem{Ma:20}
J.~Matousek, ``Point-mass method in state estimation and navigation,'' Ph.D.
  dissertation, University of West Bohemia, Czech Republic,
  10.13140/RG.2.2.30208.87044, 2020.

\bibitem{MaDuBrPaCh:23}
J.~Matou{\v s}ek, J.~Dun{\'\i}k, M.~Brandner, C.~G. Park, and Y.~Choe,
  ``Efficient point mass predictor for continuous and discrete models with
  linear dynamics,'' http://arxiv.org/abs/2302.13827, 2023.

\bibitem{MuBrBiZa:17}
C.~Musso, A.~Bresson, Y.~Bidel, N.~Zahzam, K.~Dahia, J.-M. Allard, and
  B.~Sacleux, ``Absolute gravimeter for terrain-aided navigation,'' in
  \emph{2017 20th International Conference on Information Fusion (Fusion)},
  2017, pp. 1--7.

\bibitem{Ab:10}
M.~F. Abdel-Hafez, ``The autocovariance least-squares technique for {GPS}
  measurement noise estimation,'' \emph{IEEE Transactions on Vehicular
  Technology}, vol.~59, no.~2, pp. 574--588, 2010.

\bibitem{RoJi:03}
X.~Rong~Li and V.~Jilkov, ``Survey of maneuvering target tracking. part i.
  dynamic models,'' \emph{IEEE Transactions on Aerospace and Electronic
  Systems}, vol.~39, no.~4, pp. 1333--1364, 2003.

\bibitem{NoGu:09}
P.-J. Nordlund and F.~Gustafsson, ``Marginalized particle filter for accurate
  and reliable terrain-aided navigation,'' \emph{IEEE Transactions on Aerospace
  and Electronic Systems}, vol.~45, no.~4, pp. 1385--1399, 2009.

\bibitem{DuSpa:05}
A.~Dubkov and B.~Spagnol, ``Generalized {W}iener process and {K}olmogorov's
  equation for diffusion induced by non-{G}aussian noise source, arxiv,'' 2005.

\bibitem{SiKraSo:06}
M.~\v{S}imandl, J.~Kr\'{a}lovec, and T.~S\"{o}derstr\"{o}m, ``Advanced
  point-mass method for nonlinear state estimation,'' \emph{Automatica},
  vol.~42, no.~7, pp. 1133--1145, 2006.

\bibitem{MaDuBrEl:21}
J.~Matou{\v s}ek, J.~Dun{\'\i}k, M.~Brandner, and V.~Elvira, ``Comparison of
  discrete and continuous state estimation with focus on active flux scheme,''
  in \emph{2021 IEEE 24th International Conference on Information Fusion
  (FUSION)}, 2021, pp. 1--8.

\bibitem{MaBrDu:21}
J.~Matou{\v s}ek, M.~Brandner, and J.~Dun{\'\i}k, ``Continuous nonlinear state
  prediction by finite volume method on logically rectangular grids,'' in
  \emph{2021 60th IEEE Conference on Decision and Control (CDC)}, 2021, pp.
  5890--5895.

\bibitem{SmGa:13}
V.~\v{S}m\'{i}dl and M.~Ga\v{s}perin, ``{R}ao-{B}lackwellized point mass filter
  for reliable state estimation,'' in \emph{16th International Conference on
  Information Fusion}, Istanbul, Turkey, 2013.

\bibitem{DuSoVeStHa:19}
J.~Dun\'{i}k, M.~Sot\'{a}k, M.~Vesel\'{y}, O.~Straka, and W.~J. Hawkinson,
  ``Design of {R}ao-{B}lackwellised point-mass filter with application in
  terrain aided navigation,'' \emph{IEEE Transactions on Aerospace and
  Electronic Systems}, vol.~55, no.~1, pp. 251--272, 2019.

\bibitem{DuStMaBl:22}
J.~Dun{\'\i}k, O.~Straka, J.~Matou{\v s}ek, and E.~Blasch, ``Copula-based
  convolution for fast point-mass prediction,'' \emph{Signal Processing}, vol.
  192, 2022.

\bibitem{TiStDu:22}
P.~Tichavsk\'{y}, O.~Straka, and J.~Dun\'{i}k, ``Point-mass filter with
  decomposition of transient density,'' in \emph{Proceedings of the 2022 IEEE
  International Conference on Acoustics, Speech and Signal Processing
  (ICASSP)}, 05 2022, pp. 5752--5756.

\bibitem{LiWaYaZh:19}
S.~Li, Z.~Wang, S.~S.~T. Yau, and Z.~Zhang, ``Solving high-dimensional
  nonlinear filtering problems using a tensor train decomposition method,''
  2019.

\bibitem{Sa:06}
D.~K. Salkuyeh, ``Positive integer powers of the tridiagonal {T}oeplitz
  matrices,'' in \emph{International Mathematical Forum}, vol.~22, 2006, pp.
  1061--1065.

\bibitem{St:07}
G.~Strang, \emph{Computational Science and Engineering}.\hskip 1em plus 0.5em
  minus 0.4em\relax Wellesley, MA: Wellesley-Cambridge Press, Nov. 2007.

\bibitem{RiAr:04}
B.~Ristic, S.~Arulampalam, and N.~J. Gordon, ``Beyond the kalman filter:
  Particle filters for tracking applications,'' 2004.

\bibitem{DuStMaBr:21}
J.~Dun{\'\i}k, O.~Straka, J.~Matou{\v s}ek, and M.~Brandner, ``Accurate
  density-weighted convolution for point-mass filter and predictor,''
  \emph{IEEE Trans. on Aero. and Electronic Systems}, vol.~57, no.~6, 2021.

\bibitem{DuStMa:20}
J.~Dun{\'\i}k, O.~Straka, and J.~Matou{\v s}ek, ``Reliable convolution in
  point-mass filter for a class of nonlinear models,'' in \emph{2020 IEEE 23rd
  Int. Conference on Information Fusion (FUSION)}, 2020, pp. 1--7.

\bibitem{YaDuGuDa:03}
Yang, Duraiswami, Gumerov, and Davis, ``Improved fast gauss transform and
  efficient kernel density estimation,'' in \emph{Proceedings Ninth IEEE
  International Conference on Computer Vision}, 2003, pp. 664--671 vol.1.

\bibitem{Av:22}
Avi, ``Fast binary search - matlab central file exchange,''
  https://www.mathworks.com/matlabcentral/fileexchange/30484-fast-binary-search,
  2022, accessed: September 6, 2022.

\bibitem{Lu:22}
B.~Luong, ``{FFT}-based convolution - {MATLAB Central File Exchange},''
  https://www.mathworks.com/matlabcentral/
  fileexchange/24504-fft-based-convolution, 2022, accessed: September 6, 2022.

\bibitem{Tr:22}
B.~Treeby, ``Matlab discrete trigonometric transform library - github,''
  https://github.com/ucl-bug/matlab-dtts/releases/tag/v1.1, 2022, accessed:
  September 6, 2022.

\bibitem{Br:97}
D.~H. Brainard, ``The psychophysics toolbox,'' \emph{Spatial Vision}, vol.~10,
  pp. 433--436, 1997.

\end{thebibliography}

\end{document}